\documentclass{sig-alternate}
\newcommand{\comment}[1]{}
\newcommand{\nccut}[1]{}

\usepackage{cite}
\usepackage{colortbl}
\usepackage{multirow}
\usepackage{caption}
\usepackage{subfig}
\usepackage{url}
\date{}

\begin{document}

\conferenceinfo{\copyright 
(2012) ACM. This is the author's version of the work. It is posted here by permission of 
ACM for your personal use. Not for redistribution. The definitive version was published in
KDD '12 Beijing, China, {http://doi.acm.org/10.1145/{nnnnnn.nnnnnn}}}{}
\CopyrightYear{2012} 
\crdata{978-1-4503-1462-6 /12/08} 
\clubpenalty=10000 
\widowpenalty = 10000

\title{The Untold Story of the Clones: Content-agnostic Factors that Impact YouTube Video Popularity}



\numberofauthors{5} 
\author{
\alignauthor Youmna Borghol\\
       \affaddr{NICTA \& UNSW}\\
       \affaddr{New South Wales, Australia}
\alignauthor Sebastien Ardon\\
       \affaddr{NICTA}\\
       \affaddr{Alexandria, NSW, Australia}
\alignauthor Niklas Carlsson\\
       \affaddr{Link{\"{o}}ping University}\\
       \affaddr{Link{\"{o}}ping, Sweden}
\and
\alignauthor Derek Eager\\
      \affaddr{University of Saskatchewan}\\
      \affaddr{Saskatoon, SK, Canada}
\alignauthor Anirban Mahanti\\
       \affaddr{NICTA}\\
       \affaddr{Alexandria, NSW, Australia}
}

\maketitle

\begin{abstract}

Video dissemination through sites such as YouTube
can have widespread impacts on opinions, thoughts,
and cultures.  
Not all videos will reach the
same popularity and have the same impact.
Popularity differences arise not only because of differences
in video content, but also because of other ``content-agnostic''
factors.  The latter factors are of considerable interest
but it has been difficult to accurately study them.
For example, videos uploaded by users with large social
networks may tend to be more popular because they
tend to have more interesting content, not because social
network size has a substantial direct impact on popularity.

In this paper, we develop and apply a methodology that is able
to accurately assess, both qualitatively and quantitatively,
the impacts of various content-agnostic factors on video
popularity.  When controlling for video content, we observe a
strong linear ``rich-get-richer'' behavior, with the total
number of previous views as the most important factor
except for very young videos.
The second most important factor is found to be video age.
We analyze a number of phenomena that may contribute
to rich-get-richer, including the first-mover advantage,
and search bias towards popular videos.
For young videos we find that factors other than the total
number of previous views, such as uploader characteristics and
number of keywords, become relatively more important.
Our findings also confirm that inaccurate conclusions
can be reached when not controlling for 
content.

\end{abstract}



\section{Introduction}

A vast amount of new video, audio, image, and text content
is created each year, much of it disseminated via the Internet.
What determines which items become popular and seen/heard/read by
many people, and which do not?
Although the content of the item (is it interesting, is it topical, is
it high-quality, and so on) plays an important role,
it has been widely recognized that
other ``content-agnostic'' factors can also have a substantial
impact on popularity.

For videos shared through a site such as YouTube, for example,
content-agnostic factors impacting a video's current viewing rate
include
the video uploader's social network size, the total number of previous
views to the video, the number of associated keywords, and the time
that has elapsed since the video was uploaded (the video ``age'').
Such factors can directly impact the choices of potential viewers,
as well as indirectly impact these choices through their influence on
the service provider's search and featuring algorithms.

There has been considerable
work on characterizing, modeling, and predicting the popularities
of user-generated videos.
User-generated video accesses at campus networks have been
characterized~\cite{Mahanti:07, Zink08}.
Both static and
temporal properties of view counts to a collection of user-generated 
videos have been studied~\cite{Cheng:07, Moon:09, MAY+09, Borghol:11}. 
Models have also been proposed for video popularity
evolution~\cite{Huberman:10, Moon:09, Borghol:11}. 
Many of these models are based on the
classical {\em rich-get-richer} phenomenon~\cite{Barabasi:99}, 
which suggests that a video will attract new views at a rate proportional
to the number of views already acquired.

Since both content-related and content-agnostic factors
impact popularity, however,
understanding how the various content-agnostic factors influence
popularity has been challenging.
For example, videos uploaded by users with large 
social networks may tend to become more popular because such users
generally upload
more interesting content,
not because social network size has any direct impact 
on popularity. 
Prior studies have used datasets consisting of videos with widely-varying
contents, and thus are unable to rigorously distinguish the impacts of
content-agnostic factors on popularity, from the impacts arising
from differing contents.

In this paper, we develop and apply a methodology that is able to
accurately assess the impacts various content-agnostic factors have on
video popularity.
Our methodology is based on
studying popularity differences among videos that have essentially
the same content; i.e., can be considered as ``clones'' of each other.
Popularity differences among clones can only be due to
content-agnostic factors.


Using data we collected for sets of manually-identified YouTube video clones,
we apply multivariate linear regression and other statistical methods
to systematically
determine the content-agnostic factors that most
influence a video's current popularity.
In particular, by analyzing a large number of explicit
measurable factors that are provided
through the YouTube API, we find that the most significant 
content-agnostic factors are the total number of previous views
and the video age.
We also show that determining the relative importance of
these factors without controlling for video content
(i.e., ignoring clone set memberships) would
result in inaccurate results; in particular, the relative importance of
factors such as
video age and the number of followers of the uploader would be
significantly overestimated.

When controlling for video content,
we find that ``rich-get-richer'' preferential selection
based on the total number of previous views appears to provide
a good model of video popularity evolution.
Specifically, using regression analysis we show that current video popularity,
       among videos of similar ``generation'' (age within a
       multi-year window), follows a scale-free rich-get-richer
       model with power-law exponent of approximately one.
We also show that
carrying out this analysis without controlling for
content
would result in erroneously concluding that
preferential selection is significantly weaker,
not scale-free, with a power-law 
exponent smaller than one.
We investigate
a number of possible contributors to the observed
rich-get-richer behavior,
including the ``first-mover'' advantage
and search bias towards popular videos.

The total number of previous views becomes less significant
for very young (newly-uploaded) videos that have not yet accumulated
many views.
For such videos, we show that other factors such as uploader
characteristics and the number of keywords become much more significant.
Their significance
is substantially underestimated, however, when not controlling
for video content.

The remainder of this paper is organized as follows.
Section~\ref{sec:methodology}
describes our data collection methodology and analysis
approach.
Section~\ref{sec:ImportantFactors} presents an analysis for
the relative impacts of the measured content-agnostic factors
on current video popularity,
while Section~\ref{sec:identity} shows the importance of controlling for video
content in this analysis.
Section~\ref{sec:rgr} studies the applicability of rich-get-richer
preferential selection models,
and examines contributors to rich-get-richer behavior.
Section~\ref{sec:dynamics} analyzes the content-agnostic
factors impacting the popularity of newly-uploaded videos.
Related work is discussed in
Section~\ref{sec:related}.
Section~\ref{sec:conclusions} concludes the paper.

\section{Methodology}
\label{sec:methodology}

\subsection{Data Collection}
\label{sec:data}

To analyze factors influencing video popularity, we start by identifying 
sets
of identical or nearly identical videos on YouTube. By identical 
we mean the same video content and audio soundtrack. 
We allow subtitles, variations in
encodings (quality), and small variations in video duration.
In this paper, we refer to such a set of nearly identical videos as a \emph{clone set} 
and videos in such a set as \emph{clones}. Through extensive exploration, 
search, and viewing of YouTube videos, we manually identified 48 
clone sets, each of which contain between 17 and 94 clones, 
with a median size of 29.5.\footnote{
Our dataset is available at \url{http://www.ida.liu.se/~nikca/papers/kdd12.html}.}
In total, we identified 1,761 videos.
(Our initial dataset was somewhat larger,
but we removed all videos whose duration deviated more
than 15\% from the median duration within
their clone set.)
  
%

We developed a web-based collection system that allows us  
to easily enter clone video urls into a database. Each video entered is assigned a 
clone set id and a video id. Once in the database, the system then
extracts video and uploader information using both the YouTube developer's 
API~\cite{youtubeAPI} and through HTML scraping. The system collects three 
types of information:

\begin{itemize}
\item{\bf Video statistics:} 
These include statistics such as
view count, uploader's followers count, number of comments,
``likes'' and ``favourite'' events and average rating.
For each clone set, two snapshots were 
collected, spaced one week apart.
For all videos in a clone set, 
the data collection was done 
within minutes.
 Table~\ref{table:vars} describes all variables collected.
 
\item{\bf Historical view count:}
When available, we extract historical video view counts from the YouTube HTML 
page. This information is referred to by YouTube as ``insight data''. 
We programmatically obtain this 
historical view count information by intercepting the URL request which the 
YouTube website uses to plot the view history graph. This URL contains 100 points with 
date/view count pairs. 

\item{\bf Influential events:}
The YouTube insight data also contains information on how users discover a video. 
It reveals the top 10 ``most significant'' sources of discovery, or where the video 
was linked from. Common sources of discovery include ``discovered through YouTube 
search'' and ``embedded on Facebook''. We also collect this list of referrers and, for 
each referrer, the first date of referral and the associated view count. 
\end{itemize}
The dataset used in this work was collected between February 2010 and April 2011.  

\begin{table*}[!t]	
{\tiny
\hfill{}
\begin{tabular}{|l|l|l|l|l|}
  \hline
Variable & Description & Type &  Scale & Category \\
\hline
Clone set ID & Unique clone set identifier & -- & --  & --\\ 
Capture time & Time at which this video data was captured & -- & --  & --\\
Upload time & Time at which the video was first published& -- & -- & --\\
Update time & Time at which the video was last updated& -- & -- & --\\ 
Categories count & Number of categories associated with this video& -- & -- & --\\
\hline
Next week views & Number of views between two weeks & Predicted & log & Video popularity \\
\hline
Rating average & Average rating (min and max ratings also measured) & Predictor & linear & Video popularity \\
Total comments & Number of comments & Predictor & log & Video popularity \\  
Total dislikes & Number of 'dislike' events  & Predictor & log & Video popularity \\
Total favourites & Number of time this video was 'favourited' & Predictor & log & Video popularity  \\
Total likes &Number of 'like' events  & Predictor & log & Video popularity \\ 
Total ratings & Number of ratings  & Predictor & log & Video popularity \\
Total view count &  Number of views & Predictor & log& Video popularity \\
\hline
Uploader age & Age of the uploader  & Predictor & log & Uploader characteristics \\
Uploader contacts & Number of (YouTube) 'friends' of the uploader  & Predictor & log & Uploader popularity\\
Uploader followers & Number of followers for the uploader  & Predictor & log& Uploader popularity\\
Uploader video count & Number of videos uploaded by the uploader  & Predictor & log & Uploader popularity\\ 
Uploader view count & Number of time any of the uploader's videos were viewed  & Predictor & log & Uploader popularity\\
\hline
Video age & Age of the video & Predictor & log & Video characteristics\\ 
Video keywords & Number of keywords assigned to the video & Predictor & log & Video characteristics\\
Video quality & The best quality (frame size) available for this video (higher is better)  & Predictor & linear & Video characteristics\\
\hline
\end{tabular}
}
\vspace{-8pt}
\caption{Variables collected and analyzed.}
\label{table:vars}
\vspace{-9pt}
\end{table*}



While the YouTube insight data provides valuable information
regarding a video's 
popularity evolution, it has some limitations.
First, this data is not available for all 
videos, as uploaders can choose to hide it from public view. We could 
retrieve the insight data for approximately 40\% of the videos in our dataset.
Second, the historical 
view count data includes only 100 points, irrespective of the video's age. To extract the view count at a 
specific point in time, we applied linear interpolation, which introduces an error dependent on video age.
Finally, the referrer data only reveals 10 referrers, with the exact method used by 
YouTube to select which
referrer to include in the list being unknown. 
This limits the number of views that can be mapped to a specific source, but also
leaves some uncertainty in whether there are other more significant sources not accounted for.
In our analysis we try to minimize the effect of these limitations.

\subsection{Analysis Approach}\label{sec:approach}

In this section, we introduce our analysis approach. Since our dataset contains 
multiple sets of (near) identical content, we are able to apply a range of techniques, 
both on individual clone sets and on the overall collection of videos across all clone sets.  
When using the overall collection, we can then choose to take the content (clone set id) into 
consideration or not. 
This allows us to identify factors impacting
video popularity, as well as evaluate
the errors of other methods that do not take into account the impacts of
differing video contents.  Specifically, 
we focus on the following:
 \begin{itemize}
\item {\bf Individual clone set statistics}: 
These are 
calculated for each clone set.
We present summaries of these as well as results
for example clone sets.
\item {\bf Content-based statistics}: These are calculated across all videos using an extended model
that takes into account each video's clone set identity.
\item {\bf Aggregate video statistics}: These are calculated across all videos,
ignoring clone set identity.  These statistics are used for comparison.
\end{itemize}

Our analysis focuses on the content-agnostic factors that most influence a video's
popularity, as measured by the view count over a week. To evaluate the relative
influence of different factors, we use three statistical tools. First, we characterize 
the relationships between variables using Principal Component Analysis (PCA). This
technique allows us to identify groups of variables which are responsible for different
parts of the variation. Second, we use correlation and collinearity analysis to 
identify interrelated variables, which can have a negative effect on regression results. 
For this purpose, we leverage a number of different statistical techniques, including 
pair-wise correlation matrices and auxiliary regression. Third, 
we use multi-linear regression with variable selection, to identify
a subset of the variables that captures the majority of the variations, and eliminate
variables that do not provide much information regarding popularity. 
Where appropriate, we apply standard hypothesis testing to assess the significance 
of our results. 

Several techniques used in the following assume a linear relationship between variables 
and normally distributed errors. To validate these assumptions, we first performed a 
univariate linear regression to examine the relationship between the response variable
(weekly view count) and each other variable. Second, we examine the residual plots and
corresponding tests to check that the conditions for using linear regression are satisfied.
Due to space limitations, we do not detail this preliminary analysis. We find that 
to ensure linearity with regards to the weekly view count, some variables require 
log transformation. In addition, some other variables clearly are weak predictors, 
with higher variation in their residuals.
To avoid introducing subjective biases, we did not remove such variables.
Instead, we allow the analysis to help us identify suitable candidates.
This turned out to be important as some variables are weak predictors on their
own, but complement other variables well.
The resulting variables used in the remainder of this paper and 
any transformations used are summarized in~Table~\ref{table:vars}. 


\subsection{Multi-linear Regression}\label{sec:regression}


In this section, we describe how we use multi-linear regression to determine which 
factors most influence video popularity. For this purpose, we define the response 
variable as the weekly view count (difference in view count between our two data
collections), and 
the measured factors (also called predictors) as  all the other variables. The use
of linear regression is motivated by the observed linear relationships between 
the measured predictors and the response variable. 

We perform three types of multi-linear analysis. We first use the standard multi-linear regression model
\[Y_i= \beta_0 + \sum_{p=1}^{P} X_{i,p} \beta_p + \epsilon_i, \]
where the response variable $Y_i$ is modelled as a linear function of the independent 
variables ($X_{i,p}$), and the method of least squares is used to estimate the 
coefficients $\beta_p$ for the $P$ predictors. {\em Individual clone set statistics} are obtained by 
applying the above model on each clone set independently; this allows us 
to determine which factors are the best predictors for each clone set. 
We then apply this model on all videos together, 
regardless of the clone set identity, to obtain {\em aggregate clone set statistics}. 
This allows us to evaluate the error when not using our content-aware approach as
discussed below.

In order to obtain {\em content-based statistics}, we design an extended model 
that incorporates a categorical variable for 
the clone set identity. This model is useful in understanding the influence of individual 
clone sets on the regression, and whether or not the classification makes a difference. 
Assuming that we have $K$ clone sets (or categories), we introduce $K-1$ 
additional category variables, each capturing the relative difference against a reference
clone set.  The extended multi-linear model is then given as:
\[ Y_i = \beta_0 + \sum_{p=1}^{P} X_{i,p} \beta_p + \sum_{k=2}^{K} {Z_{i,k} \gamma_{k}} + \epsilon_i, \]
where $K$ is the number of clone sets; $P$ is the number of predictors;
and $Z_{i,k}$ is the category regressor,
encoded as $Z_{i,k}= 1$ if clone $i$ is from clone set 
$k$, and as 0 otherwise. Note that $\gamma_{k}$ can be
interpreted as the relative distance between the
regression lines of clone sets 1 and $k$, or in other words,
a measure of their relative popularity.

\section{Factor Strength}
\label{sec:ImportantFactors}

\nccut{
In this section, we present our factor strength and importance analysis and results. 
We first describe some preliminary analysis which qualitatively explains the variation in 
weekly view count from the variables, then present our regression model, 
variable selection, and reduced models. 
}

\subsection{Preliminary Analysis}
\label{sec:preliminary}

\begin{figure}[t]
\centering
\includegraphics[trim = 0mm 8mm 0mm 0mm, width=0.45\textwidth]{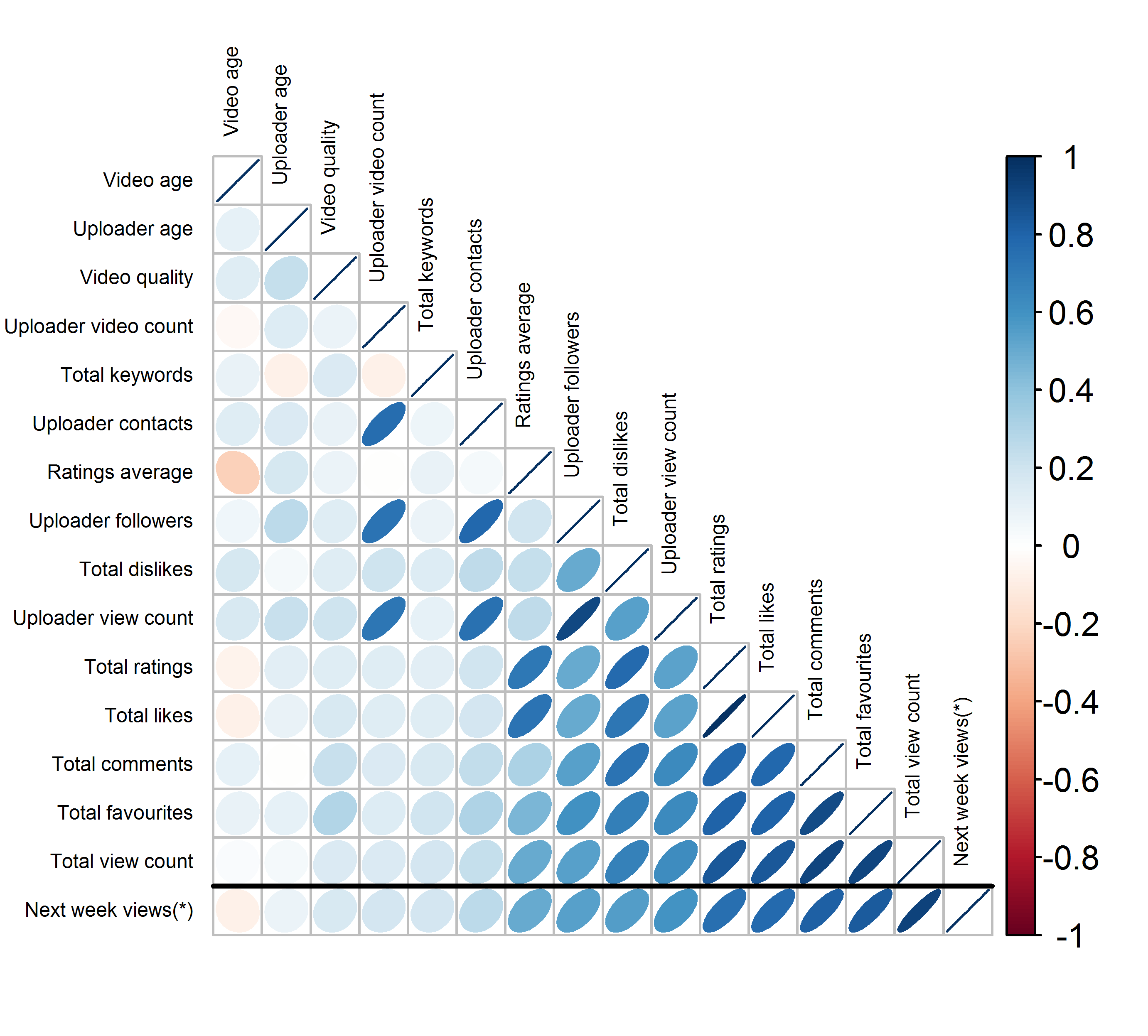} 
\vspace{-2pt}
\caption{Correlation matrix for clone set 41.}
\label{fig:correlation}
\centering
\includegraphics[trim = 0mm 4mm 0mm 0mm, width=0.4\textwidth]{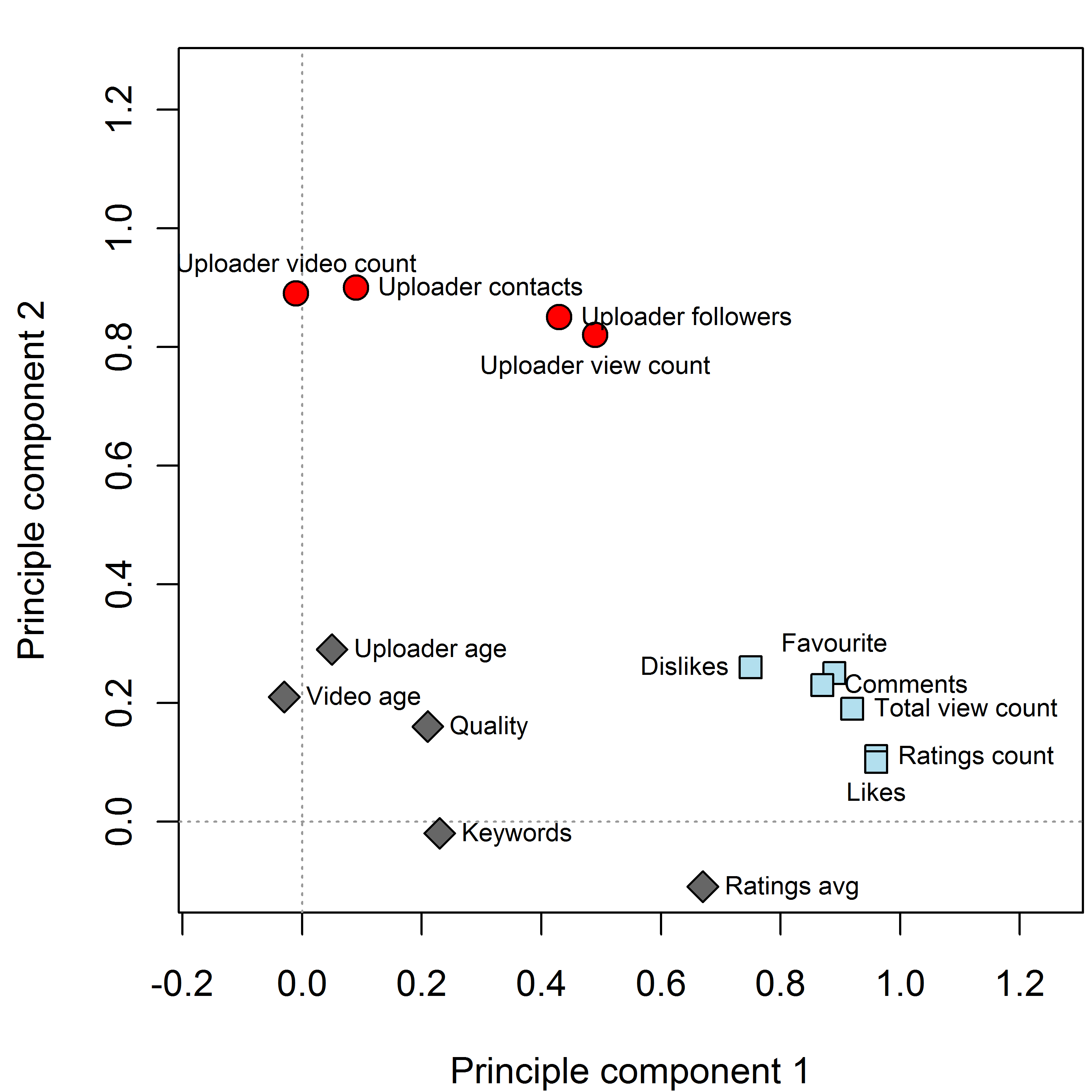}
\vspace{-2pt}
\caption{Principal components plot for clone set 41.}
\label{fig:PCA}
\vspace{-10pt}
\end{figure}

Before looking at which factors best capture the future popularity, we perform a preliminary
analysis to discover any correlations between the factors themselves, and if there are
groups of variables that provide redundant information and/or explain the same variation.
First, we investigate the strength of the linear relationships among the  variables using Pearson's correlation.  
Figure~\ref{fig:correlation} shows the correlation matrix plot for 
an example clone set. The variables in the matrix plot are ordered based on their correlation with the response variable,
and each entry shows the pairwise correlations between the corresponding two variables.
The correlation's magnitude is represented by the ellipse symbol,  and its sign is represented by colors (and slope), 
with red (with slope to the left) used for negative values and blue (with slope to the right) 
for positive values. We note that many of the variables have high pairwise correlation 
and very  similar clustering of the pairwise correlation for most clone sets.
In particular, two sets can be  identified: 
(i) the set of variables related to the past video popularity
(i.e., the total view count, favourite count, comment count, ratings count, likes and dislikes), and
(ii) the set of variables related to the uploader characteristics  
(e.g., the number of uploader followers, contacts, videos, and views).

Second, we apply Principal Component Analysis (PCA) on each of the individual clone sets.
We  find the two aforementioned variable groups  to correspond to the two primary 
principal components, particularly for younger clone sets.  
Figure~\ref{fig:PCA} shows a PCA plot of the first two principal components for one such clone set.
Referring to Table~\ref{table:vars}, these roughly correspond to {\it video popularity} 
and {\it uploader popularity} metrics, respectively.  For many other clone 
sets, particularly clone sets with big variation in video age, 
other {\it video characteristics} (such as video age and video quality) forms 
a third important component.

 
Understanding and detecting collinearity is important, 
as interrelated variables could 
negatively affect the regression results.
To find out whether predictor $X_i$ is a linear combination of
other predictors, we run auxiliary regressions
to determine the coefficient of determination $R^2_i$ 
of how well the remaining explanatory variables $X_{j \neq i}$
explain $X_i$.

While the full results are omitted due to lack of space, 
the auxiliary regressions on all individual clone sets
and on all videos aggregated across clone sets show that the $R^2_i$ values of the 
following regressor factors exceed the overall $R^2$ value of the model 
including all the explanatory variables: the total view count, the number of 
times a video was favourited, the number of comments, the number of ratings, 
the number of times the video was ``liked'' or ``disliked'', as well as the total number 
of views to all videos uploaded by the uploader, and the count of the uploader's followers.
We note that these factors fall into the two previously identified groups 
of correlated variables. Overall, these results provide evidence of a serious 
collinearity and its sources. 


\subsection{Variable Selection within Clone Sets}\label{sec:intra-clone set}

To determine which 
predictors
have the most impact on the popularity of a 
clone within a clone set, we applied multivariate regression analysis on individual clone sets. 

\begin{figure}[!t]
\centering
\includegraphics[width=0.42\textwidth]{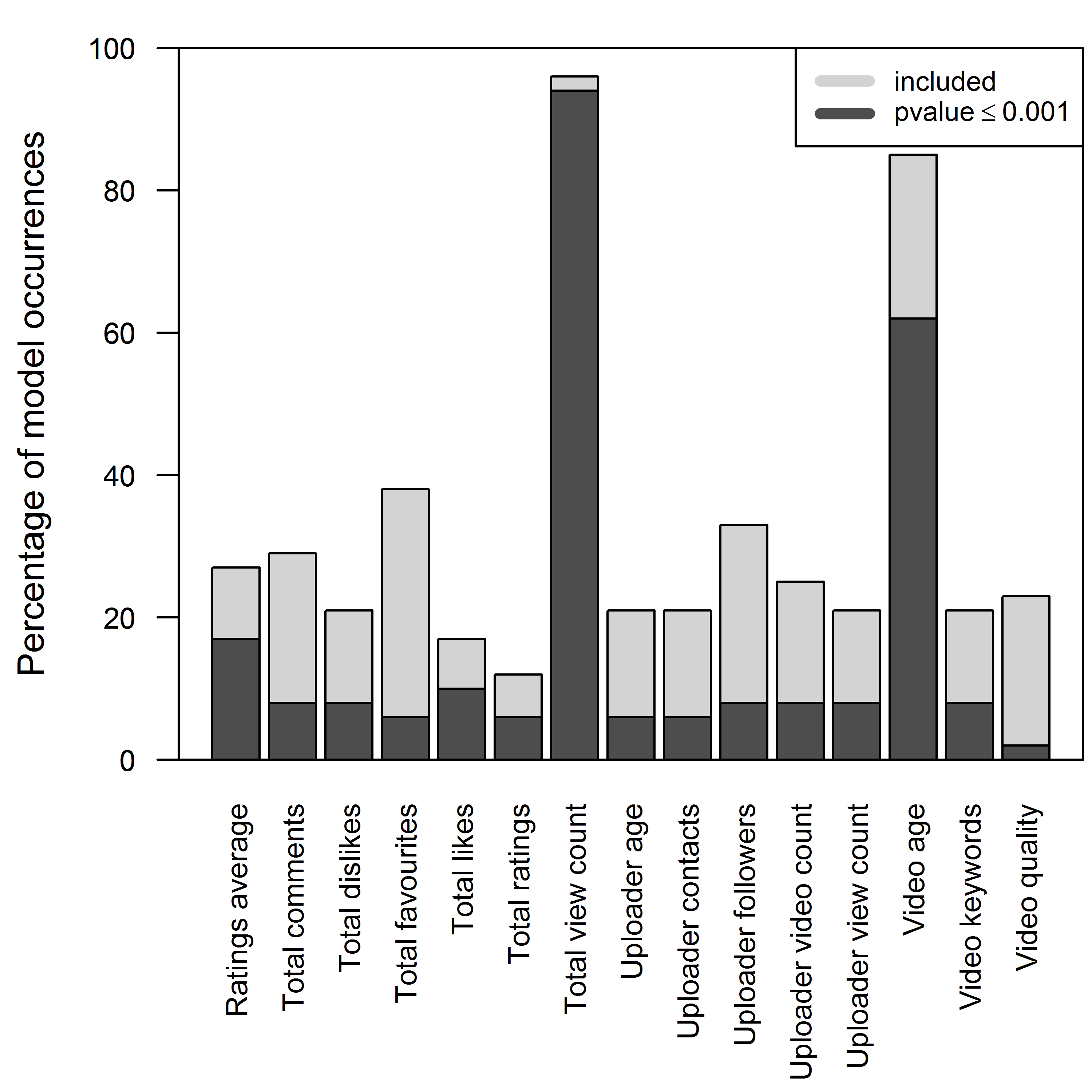}
\vspace{-2pt}
\caption{Percentage of occurrences in the set of ``best models'',
using the best subset approach with Mallow's $C_p$.
Dark color shows fraction of models in which the variable was 
selected while having a p-value smaller than 0.001 in the final model.
In the remaining occurrences the variable was selected, but with a higher p-value.}
\label{fig:occurrence}
\vspace{-8pt}
\end{figure}


To limit the impact of collinearity and additional noise on the regression models, 
we  eliminate redundant variables using the best subset search technique and
Mallow's $C_p$ as the selection criterion~\cite{AllenStats}.
We have obtained qualitatively similar results using other commonly used regression methods 
and selection criteria. The results in Figure~\ref{fig:occurrence} show that the total view count is the most important 
explanatory variable.  It is selected in 92\% of the total set of ``best models'', 
and is determined to be highly significant. The video age is the second most important 
predictor, being very significant and 
being the second most common variable in the models. 
While the video age did not appear 
to be
a good predictor on its own, as exemplified by the 
ordering in Figure~\ref{fig:correlation} 
and low individual $R^2$ values (with a median of 0.081), 
its frequent inclusion indicates that it 
accounts for different variations than the total view count. 
This has also been observed in some of our PCA analyses.
It is also interesting to note that other independent variables, such as variables related to 
the uploader characteristics, did not appear important in the original regression are now 
significant when selected in the final model, and are often significant for younger clone sets. 
One factor that seldomly is significant (even when included) is the video quality.  
In part, this may be a consequence of use of default
        encodings.  However, although our analysis does
        not find a significant linear relationship, we
        believe that quality differences may be important in clone
        sets with wide variations in video age and associated
        wide variations in quality.  The quality
        variations in such cases may play a 
        role in making age our second most important predictor.

From Figure~\ref{fig:occurrence}, we can see that the best subset approach with
Mallow's $C_p$, on average, reduces the number of variables by about 60\%. 
The multiple $R^2$ 
values for the chosen models are then only slightly smaller than the original  
$R^2$ value of the full model.

An interesting observation is that the most influential factors also 
are the only statistics available to the YouTube users, when searching for a video.


\section{Impact of Content Identity}\label{sec:identity}

The regression analysis presented in the preceding section was applied on individual 
clone sets. Using the model extension presented in Section~\ref{sec:regression}, 
we perform regression analysis over the entire dataset while taking into account 
the content identity, and thus by extension study the impact of the video content 
on popularity dynamics.  Evaluating the importance of the clone set categorical variable 
is important as it allows us to separate the impact of content-related and content-agnostic factors. 

\begin{table}[!t]
\begin{center}
{\tiny
\begin{tabular}{|r|r|r|r|r|}
  \hline
 & Estimate & Std. Error & t-value & p-value \\ 
  \hline
Total view count ($\beta_i$) & 1.100 & 0.013 & 87.83 & 0.000 $***$  \\
Video age ($\beta_i$) & -1.008 & 0.039 & -25.80 & 0.000 $***$  \\
Clone set (min$_k \gamma_k$) & -0.727 & 0.348 & -2.08 & 0.037 $~~~~*$ \\
Clone set (max$_k \gamma_k$) &  2.802 & 0.345 & 8.08 & 0.000 $***$ \\
   \hline
\end{tabular}
}
\end{center}
\vspace{-10pt}
\caption{Summary of extended regression analysis using categorical variables
for clone set identification.  With 95\% confidence, the rejection rate of the hypothesis 
that the category variables ($\gamma_k$) are equal to zero is 94\%.}  
\label{table:dummy}
\raggedright
{\tiny
\begin{tabular}{|c|c|c|c|c|c|}
\hline
& View count &  + age    & + followers                    &  all                  \\
& (1 variable)        &  (2 var.) &  (3 var.)         & (15 var.)               \\
\hline
Individual (mean)    &  0.788 &  0.864  & 0.871 & 0.933 \\
Individual (median)   &  0.803 &  0.873  & 0.875 & 0.940  \\
Individual (41) & 0.861  & 0.870  & 0.874  & 0.895\\
\hline
Content-based  & 0.792  &  0.850  & 0.852 & 0.855\\
\hline
Aggregate  & 0.707 &  0.808  & 0.808  & 0.821\\
\hline
\end{tabular}
}
\caption{Summary of $R^2$ values for example models.}
\label{table:rsquare}
\vspace{-14pt}
\end{table}

We perform the content-based regression analysis using the most important explanatory variables
identified in Section~\ref{sec:intra-clone set}. We use our default clone set ordering, where sets are 
numbered from 1 to 48, and choose one clone set as the baseline set.  

Summary results are presented in Table~\ref{table:dummy}, for the baseline clone set number 1. 
The coefficients of the category variables ($\gamma_k$) explain by how 
much the intercept of the selected clone set differs from the intercept 
of the baseline clone set. The significance of the categorization, i.e., the impact of video content, 
is then measured by the corresponding p-values. We also report range values 
min$_k \gamma_k$ and max$_k \gamma_k$, across the 47 non-baseline clone sets.

We find that 44 out of 47 category variables have p-value smaller than 0.05.
When averaging over all possible baseline clone sets, we found approximately 
60\% of the category variables to be significant. This illustrates the importance of 
taking clone identity into consideration.

As a second step to evaluate the importance of video content, we compare the regression 
analysis results of the content-aware extended model and the regular individual clone set 
models with the aggregate model which ignores clone identity.

For each model type, we used four different models: three partial models and one full model.
The first partial model includes only the view count variable, the second model includes 
both the view count and the video age, and the third model further adds the uploader 
followers.

Table~\ref{table:rsquare} shows the coefficient of determination $R^2$ 
for each model when running the regression analysis on each clone set individually (``Individual''), across 
all clones and clone sets as an aggregate (``Aggregate''), and when we take the clone identity into account 
using category variables (``Content-based''). 
Comparing the last two rows, we note that the ``Content-based'' 
models consistently explain a larger portion of the variation, as evidenced by higher $R^2$ values.
This is another indication that taking into account the clone identity is important in modelling 
video popularity. 


Table~\ref{table:rsquare} also reveals that the view count by itself explains
the biggest percentage of the variance, especially when taking into account
the clone identity.
Adding the video age variable increases the $R^2$ values
relatively significantly. Adding the uploader followers variable can
result in an occasional incremental increase in the goodness of fit,
while the other variables impact is even less important.

However, perhaps more importantly, this table also shows that 
if one tried to analyze the relative importance of age, followers, etc., without controlling for video content, 
one would conclude that factors such as age and followers are relatively more important (compared to view count) 
than they really are. This is illustrated by comparing the difference in values from left to right, for the
aggregate and the content-based models.  
In one case, $R^2$ is improved by 0.114, and in the other by only 0.063.

The next section  will take a closer look at the impact clone identity 
may have on predictive models,
such as the rich-get-richer model.

\section{Rich-get-Richer}
\label{sec:rgr}

Prior works have suggested that video popularity evolves according to rich-gets-richer preferential selection~\cite{Barabasi:99}
or a variant thereof (e.g.,~\cite{Huberman:10, Moon:09}), wherein the current viewing rate of a video
is proportional to the total number of views the video has already acquired.
In Section~\ref{sec:rgr:models}, we evaluate whether or not our data is
consistent with a rich-get-richer model of popularity evolution.
Section~\ref{sec:rgr:fma} considers 
a more restricted form of rich-get-richer behavior,
the ``first mover'' advantage.
Finally, Section~\ref{sec:rgr:vdf} explores other phenomena that
may result in rich-get-richer behavior, including search bias
towards popular videos.

\subsection{Models}
\label{sec:rgr:models}

We consider rich-get-richer models wherein
the probability $\Pi(v_i)$
that a video $i$ with $v_i$ views will be selected for viewing
follows a power law
\[ \Pi(v_i) \propto v^{\alpha}, \]
where $\alpha$ is the power exponent.

Perhaps most interesting is when $\alpha=1$.
This case corresponds to linear preferential selection, 
was considered by Barabasi and Albert~\cite{Barabasi:99},
and can be shown to result in a scale-free
distribution (in our context, of total view counts).

Basic rich-get-richer models as described above
        consider only the number of accumulated views as
        a determinant of the rate of acquiring additional
        views.  With YouTube videos, however, user interest
        in particular subjects changes over time, causing
        a deviation from rich-get-richer behavior when one
        considers a collection of such videos with
        differing contents.  An important question is whether
        a rich-get-richer model is applicable when one
        removes the impact of changing user interests,
        as we are able to do with our clone-based methodology.


\begin{table}[!t]
\begin{center}
\begin{tabular}{|l|l|l|l|}
\hline
\multicolumn{1}{|c|}{} & \multicolumn{1}{|c|}{Slope estimate} &
\multicolumn{2}{|c|}{Confidence intervals} \\
\hline
Metric        & $\alpha~(\sigma)$ & 90\% & 95\% \\
\hline
Individual    &  1.027 (0.091) & 0.988-1.065 & 0.981-1.073  \\
\hline
Content-based &  1.003 (0.014) & 0.98-1.027 & 0.976-1.031 \\
\hline
Aggregate  &  0.932 (0.016) & 0.906-0.958 & 0.901-0.963 \\
\hline
\end{tabular}
\vspace{-6pt}
\caption{Rich-get-richer slope estimates.}
\label{table:slope-hyp}
\vspace{-16pt}
\end{center}
\end{table}


To answer this question, we first identify within each
        clone set videos of similar ``generation'' (age
        within a multi-year window).  We restrict attention
        to videos of similar generation to avoid our analysis
        being impacted by wide variations in video quality
        (or other generation-related effects).
        Specifically, for each clone set, we first find the
        video clone with the highest current popularity (i.e.,
        the video that acquired the most additional views
        during our one week measurement window).  We
        then consider only the videos in the clone set that
        were uploaded within two years of the upload time of
        this video.

We now take a closer look at the impact differences in video identity
can have on the rich-get-richer phenomena.  We examine how the rate at which videos attain new
views depends on the total view count using univariate linear regression (using log-transformed data).
All three analysis approaches, namely regression analysis on individual clone sets, on the aggregate, and 
the aggregate considering content identity, were applied. 
Table~\ref{table:slope-hyp} summarizes our results.

The first column in Table~\ref{table:slope-hyp} shows the coefficient estimates and standard
deviation resulting from the univariate regression analysis.
The second and third columns show the corresponding confidence intervals.
For the individual clone sets, and the extended content-based model, $\alpha$ is typically
equal or slightly higher than one.  The selection rate is linearly dependent on the current total view count,
suggesting
that the popularity evolution is scale free,
and strongly controlled by rich-get-richer behavior.
For the aggregate model, $\alpha$ is less than one, indicating 
popularity evolution
that could result in a much more
even popularity distribution than that suggested by the pure (linear)
rich-get-richer dynamics.

\subsection{First Mover Advantage}
\label{sec:rgr:fma}

Rich-get-richer behavior may result in part from a ``first mover'' 
advantage. The first video to include particular content may have 
already achieved significant dissemination by the time that clones appear,
causing it to acquire new views at a higher rate (for example, via 
recommendations from previous viewers, featuring, or bias in search algorithms). 
Using our clone-based methodology, we now evaluate the advantage of being 
the first to upload particular content.

To track video popularity over time, we use YouTube's insight data collected through HTML
scraping. As a first step, we consider the success of the first mover in each cloneset, where
a success event for a particular video is defined as when that video accumulates the largest
number of total views compared to all other videos within the clone set.
We first consider how often the most successful video within a clone set is the first to 
either be uploaded or discovered through search.
Table~\ref{tab:firstMover} shows the number of times the video clone
that obtained the highest total view count was first, second, third,
fourth, or fifth, among the
videos in the clone set, to be 
uploaded or found through search.
Overall, the winner was uploaded first among
the videos in the clone set in 27.1\% of the observed cases, and was 
among the first five in 60.4\% of the cases. 
Similarly, the winner is the first to be found through search in 66.7\% of the cases 
for which we have (insight data) statistics, and among the first five to be found 
through search in 92\% of the cases. Clearly, there is a significant advantage to being first mover.

\begin{table}[!t]
\centering
\begin{tabular}{|l|c|c|c|c|c|c|}
  \hline
 & 1st & 2nd & 3rd & 4th & 5th & later\\
\hline
Winner uploaded & 27.1  & 12.5    & 8.3   & 6.3   & 6.3   & 39.6\\
Winner searched & 66.7 & 8.3   & 0.0   & 8.3   & 8.3   & 8.3\\
 \hline
\end{tabular}
\vspace{-6pt}
\caption{The percentage of times a video clone that obtained the highest total view count
was the first, second, third, fourth, fifth (or later) among the  videos in the clone set
with respect to being uploaded or searched.  (Clone sets with relevant statistics considered.)}
\label{tab:firstMover}
\vspace{-12pt}
\end{table}


While these results suggest that the first mover typically is
relatively successful, it is interesting to note that there are
cases where other videos have been able to surpass the first mover
in popularity.
What is it that allows some other video to overtake 
the spot as the most popular clone?
Section~\ref{sec:dynamics} 
takes a closer look at some influences 
that can cause such overtakings.


\subsection{Video Discovery and Featuring}
\label{sec:rgr:vdf}

We now examine the roles that video discovery and featuring mechanisms
may play in the observed rich-get-richer preferential selection behavior.
Aspects such as featuring on YouTube, ranking of a video in
YouTube search, and embedding of a video on external sites,
are difficult to capture over time.
Nonetheless,
the ``video referrers'' part of the YouTube insight data provides
(for some videos) additional information necessary for our analyses.
The results presented here are based on analysis of clone sets
that have multiple videos with insight data.
We use YouTube's classification of registered
referrers.

\begin{figure}[!t]
\centering
\includegraphics[width=0.375\textwidth]{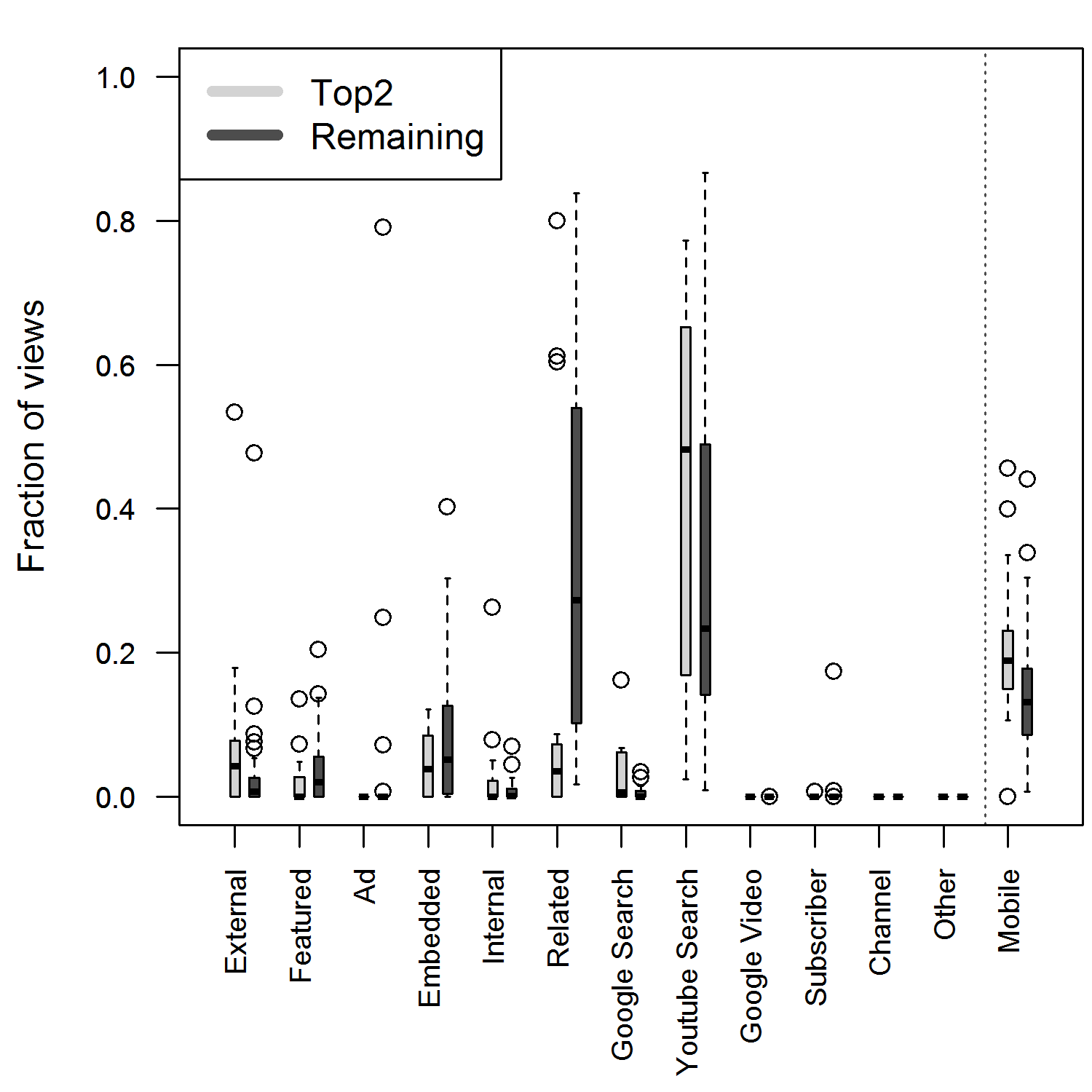}
\vspace{-10pt}
\caption{Boxplot of the average fraction of views (per cloneset) coming through different referrer categories.}
\label{fig:boxplot-fracviews-cloneset}
\vspace{-6pt}
\end{figure}

We first consider how the most popular videos within each clone set have obtained their views,
compared to their less popular counterparts.
Figure~\ref{fig:boxplot-fracviews-cloneset}
compares the average fraction of views coming through
different referrer types for the most popular clones, with that of 
the remaining clones.\footnote{We present the mobile referrers separately
as it is not a source of discovery per se, but it nevertheless impacts discoverability, as more
users are accessing videos exclusively through mobile devices.}


The results are somewhat counter-intuitive. Notice that the ``Top 2'' most popular 
videos are not necessarily the videos that are 
prominently featured or externally linked. Instead, the search discovery method alone accounts for most of 
the difference. For example, for the search referrer category, the median of the top clones is almost equal 
to the 90th quartile of the remaining videos. 
The less successful clones get most of 
their views through related (video) referrals.

\begin{figure}[!t]
\centering
\includegraphics[trim = 0mm 8mm 0mm 0mm, width=0.375\textwidth]{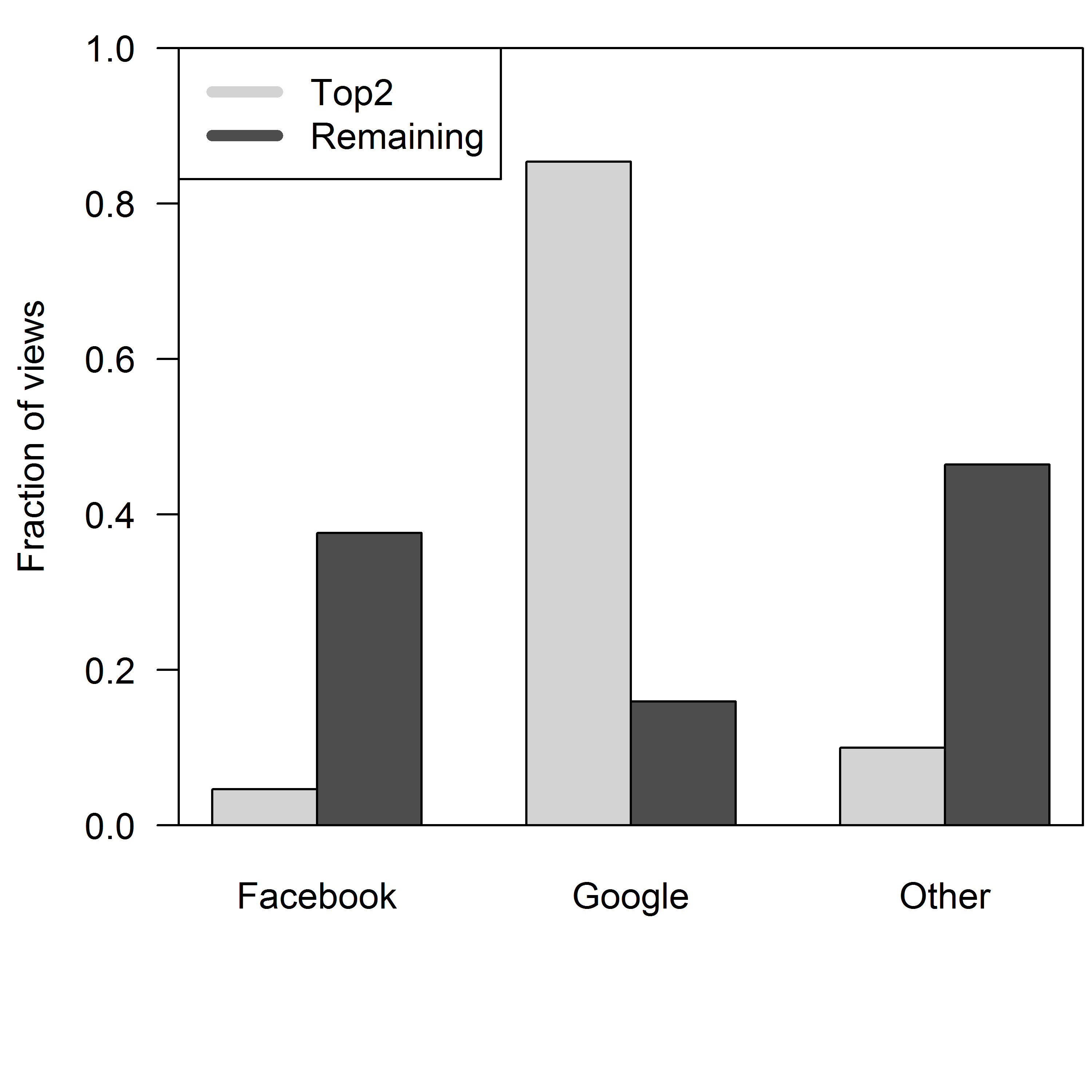}\\
\vspace{-16pt}
\caption{The fraction of views coming through external sources, for clones that are externally linked.}
\label{fig:bar-fracviews-extra}
\vspace{-6pt}
\end{figure}

Figure~\ref{fig:bar-fracviews-extra} shows the fraction of views coming through external
referrers only (not including embeds). Note that Google is shown as an external
source of traffic and it is  driving most of the external views of the
popular clones.  Google is considered an external referrer because 
views may come from a number of Google non-search services such as 
Google News, Google Reader, and Google Group posts.

\begin{figure*}[!t]
\begin{center}
\begin{tabular}{cc}
\includegraphics[width=0.35\textwidth]{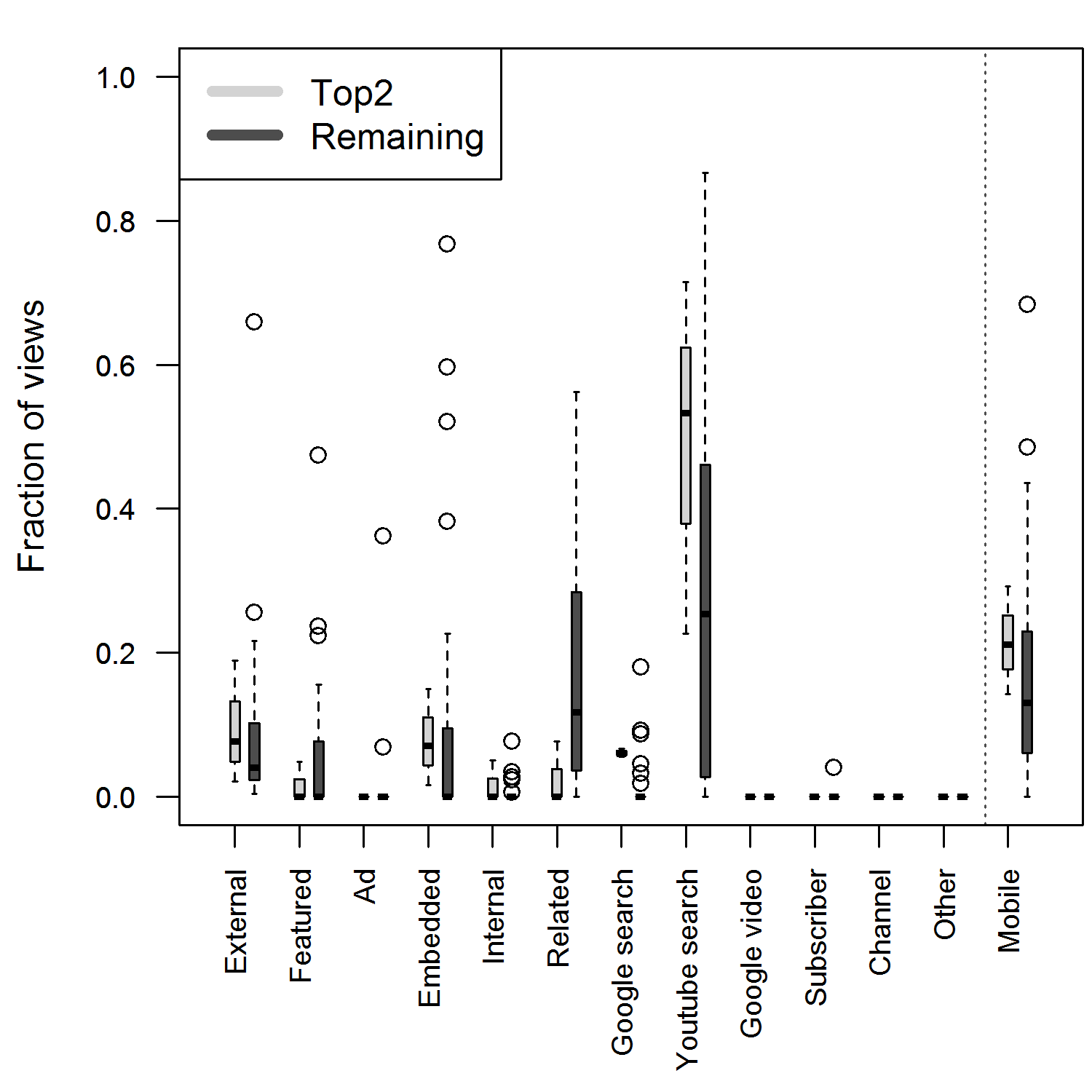}&
\includegraphics[width=0.35\textwidth]{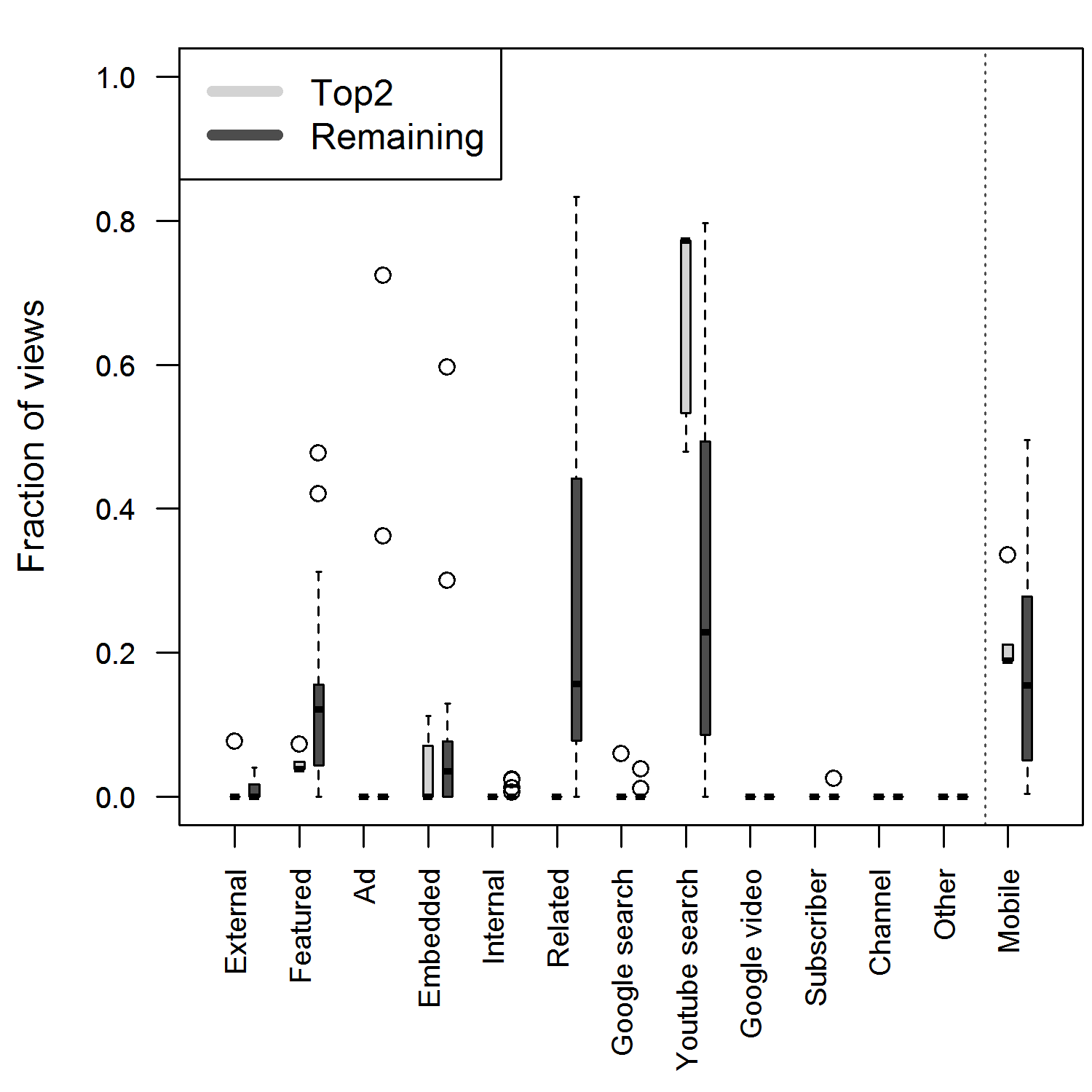}\\
(a) Externally linked clones &
(b) Featured clones\\
\end{tabular}
\vspace{-10pt}
\caption{Boxplot of the fraction of views of clones externally linked and featured, 
coming through different referrer categories.}
\label{fig:boxplot-fracviews-clone-extra}
\end{center}
\vspace{-10pt}
\end{figure*}

Overall, the highest fraction of clicks to a video is coming through the search referrers.
As all videos can potentially be found through search, but not all videos are featured 
or embedded on external websites, we take a closer look at these referrers. 
Figures~\ref{fig:boxplot-fracviews-clone-extra}(a) and~\ref{fig:boxplot-fracviews-clone-extra}(b)
show the corresponding boxplots of the fraction of views coming through different referrer types
for only the clones that are featured and externally linked, respectively.
The same conclusion applies to these data subsets.  Search referrers are the most powerful 
in terms of the percentage of traffic they bring. Further, the biggest differences between 
successful and less successful videos can be seen in the fraction of views owing to search 
and mobile referrers.

Recall that we are considering multiple videos containing essentially
the same content, and this allows us to remove biases
introduced because of differences in content (e.g., popular
content is more likely to be searched than non-popular content).  
Our results suggest that successful videos are much more 
prominently selected through searches.
This could 
occur because of 
YouTube's internal search mechanism,
the keywords associated with the videos,
the keywords entered by the users,
user biases when selecting among search results,
or a combination thereof.
For example,
people may be more likely to pick the first search results
than pick items lower down on the list, or to pick videos with
higher view counts (visible to the user at the time of selection)~\cite{SaWa08}.

As previously discussed, the first mover advantage can be important 
for the success of a video.
In addition to being uploaded, it is important that the video is discovered 
and/or made available through different paths.  Using correlation analysis, 
we have observed that there is often a significant positive correlation 
between the total view count and the order in which clones are first referred, 
featured, or accessed through mobile devices.  While omitted, these results suggest 
that there also is a \textit{first-discovery advantage}, where videos discovered 
earlier through internal search methods, featured earlier, or that are accessed through 
mobiles earlier, tend to be ranked higher.

\section{Factors Impacting Initial Popularity}\label{sec:dynamics}

This section considers factors impacting the view count early in a video's life, 
which in turns impact the overall video popularity due to the 
rich-get-richer behavior, as shown in the previous section. 

\begin{figure}[!t]
\includegraphics[trim = 0mm 12mm 0mm 18mm, clip, width=0.45\textwidth]{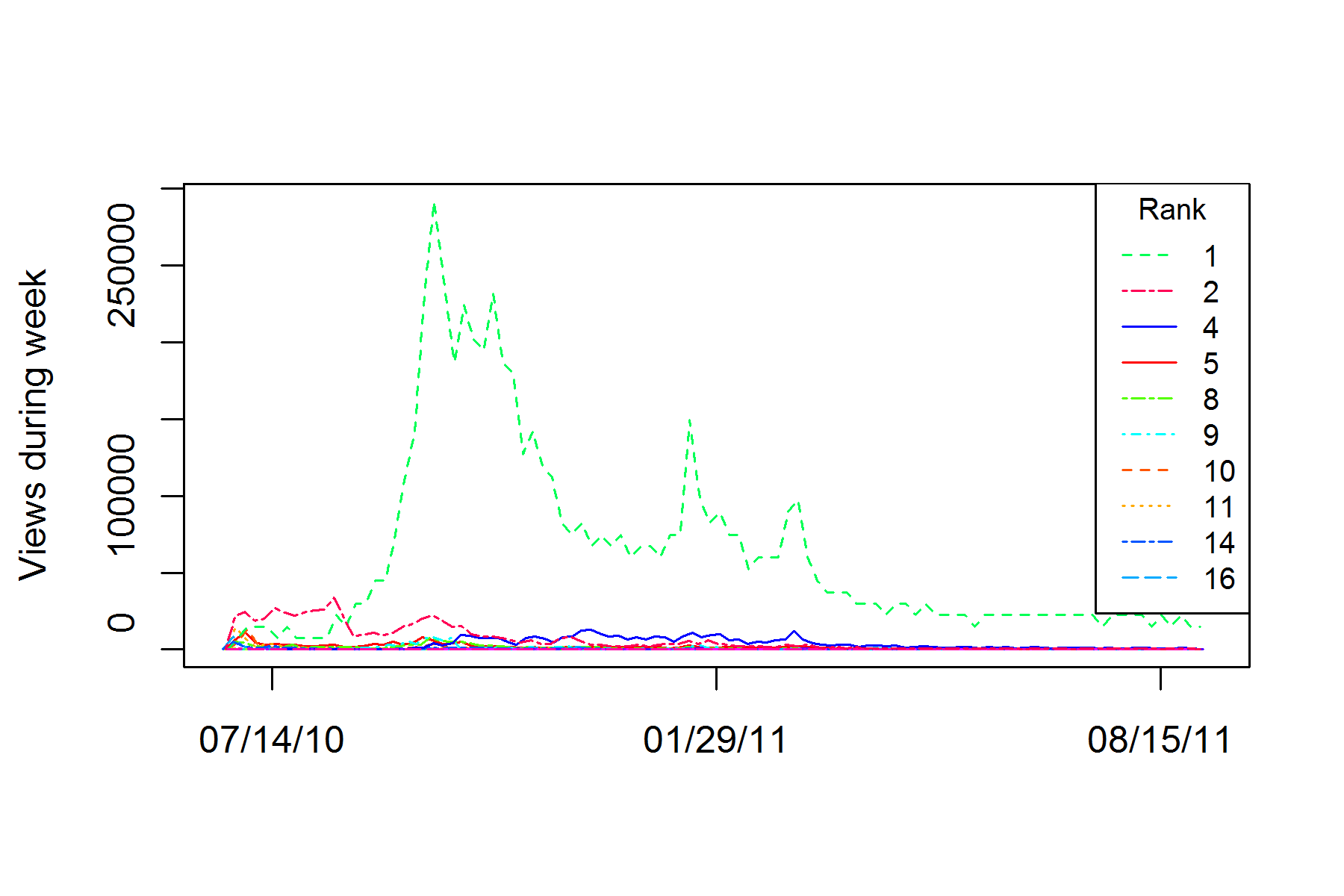}
\vspace{-10pt}
\caption{The weekly views for a number of example videos in clone set 14 (18 clones).}
\label{fig:evolution-38}
\vspace{-6pt}
\end{figure}

\subsection{Uploader Characteristics}

We analyzed the YouTube social network size of the uploaders observed in our dataset. 
In general,  uploaders of top-ranked videos have large social networks. 
Furthermore, manual examination of the top uploaders confirmed
that they are often commercial entities.
Usually commercial uploaders are the ``first-movers''.
In fact, even when they were not, it appears that their videos often manage to move ahead in popularity.
Figure~\ref{fig:evolution-38} shows an example clone set where a commercial user (video with rank 1) 
catches up and surpasses a private uploader (video with rank 2) even though the former 
was not the first uploader. 
This is a typical example of the impact of uploader
characteristics on the popularity.

\subsection{Age-based Analysis}

While the total view count can be an important factor when
predicting a video's future popularity, it is less useful
for young videos and not useful at all for a newly uploaded
video which starts with a view count of zero.  For such
videos, other predictors such as the size of the uploader
social network and the prior success of the uploader may
be important, as seen in Section 6.1 and in our
PCA analysis in Section 3.  We now perform an age-based
regression analysis to determine how the relative importance
of the total view count changes with time, relative to these
more static factors.

\begin{table}[!t]
{\tiny
\begin{tabular}{|l|c|c|c|c|c|c|c|c|}
\hline
 & \multicolumn{4}{c|}{Aggregate} & \multicolumn{4}{c|}{Content-based} \\
\hline
Predictor / Age & 1d & 3d & 7d & 14d & 1d & 3d & 7d & 14d \\\hline
View count & 0.44 & 0.42 & 0.50 & 0.55 & 0.60 & 0.59 & 0.66 & 0.70\\\hline
Video quality &  \multicolumn{4}{c|}{0.08} & \multicolumn{4}{c|}{0.35}\\\hline
Number of keywords & \multicolumn{4}{c|}{0.04} & \multicolumn{4}{c|}{0.36}\\\hline
Uploader view count & \multicolumn{4}{c|}{0.41} & \multicolumn{4}{c|}{0.64}\\\hline
Uploader followers & \multicolumn{4}{c|}{0.40} & \multicolumn{4}{c|}{0.58}\\\hline
Uploader contacts & \multicolumn{4}{c|}{0.19} & \multicolumn{4}{c|}{0.42}\\\hline
Uploader video count & \multicolumn{4}{c|}{0.08} & \multicolumn{4}{c|}{0.38}\\\hline
Uploader age &  \multicolumn{4}{c|}{0.02} & \multicolumn{4}{c|}{0.35}\\
 \hline
\end{tabular}
}
\caption{Age effect on $R^2$ values when taking into account 
clone set identity (content-based) and when not (aggregate).}
\label{table:r2-age-clone}
\vspace{-8pt}
\end{table}

Table~\ref{table:r2-age-clone} shows the coefficient of determination $R^2$
between the predictors in the first two weeks since a video's upload
and the total view count at the half-year point since upload.
We calculate the total view count of videos at
1 day, 3 days, 1 week, 2 weeks, and half a year using 
the historical view statistics. Linear interpolation
is needed to calculate the approximate total view count at specific time thresholds,
as the data provides only 100 points, equally spaced through the video's lifetime.
The file-related information and the uploader characteristics properties are assumed constant.
The first four columns show results for the aggregate set of videos and
the last four columns show results when clone set identity is accounted for.

These results show that the total view count quickly becomes the strongest predictor
of the view count at the half-year point.
The results also confirm that during the early stages of a video's lifetime, 
the uploader's social network is a more significant factor than the total view count.
Indeed, already at upload, approximately 64\% of the variation in views can be
 explained by the uploader view count alone, and it takes a week for the total view 
 count to become a similar or better predictor than the uploader social network.
The impact of the uploader characteristics are significant in the
beginning, probably because an established social network
is a source of initial views from subscribers.

Finally, we note that some factors have much more impact when the influence 
of the content is considered through the clone set identity factor.  
For instance, factors such as the number of keywords
and the video quality,
have a great impact in the early
stages of a video's lifetime. The keyword metric, although appearing to 
be insignificant in the aggregate analysis, is an important factor when
a video is first uploaded, explaining up to 36\% of the variation in views.
This may suggest that keywords in fact may be one of the main
factors in helping find the video in the first place
(when competing against videos with the same content).
The more targeted keywords a video has, the greater the probability
that it will be discovered after its upload. 

\section{Related Work}
\label{sec:related}

To the best of our knowledge, no prior work has separated out the
impacts of content-related and content-agnostic factors on video
popularity.  However, there has been considerable prior work concerning
measurements, analyses of these measurements, and/or models, for
various user-generated video properties including popularity.

Many prior works have analyzed different aspects of
user-generated video metrics
such as total views, total ratings, total comments, and uploader 
social network size (e.g., see~\cite{Moon:09, MAY+09, Cheng:07}).
For instance, 
Mitra et al.~\cite{MAY+09} compared four video sharing workloads
and established the presence of ``invariants'' among their
characteristics, such as heavy-tailed total view count
distributions and positive correlation between total views
and total ratings to a video.  

Cha et al.~\cite{Moon:09} postulated that heavy-tailed view
counts can be explained by combining
a ``rich-get-richer'' model~\cite{Barabasi:99} 
with a limited fetch model.
Szabo and Huberman~\cite{Huberman:10} found that the total views received 
soon after a video is uploaded provides a strong indication of
its total future view count; they
leveraged this observation to develop a prediction model for video views.  
More recently, Borghol et al. \cite{Borghol:11} empirically
demonstrated that individual video popularity is highly unstable
and unpredictable, and 
proposed a model for how the popularity
statistics of a collection of recently-uploaded videos evolve over time,
instead of considering popularity evolution for individual videos.

YouTube's internal search and recommendation engines
have been noted to be an important source
of video views~\cite{Zhou:10, Figueiredo:11}.
For instance, Zhou et al.~\cite{Zhou:10} observed 
strong correlation between the total view count of a video and
the view count of its ``related'' videos.
Similarly, Figueiredo et al.~\cite{Figueiredo:11} noted
that search and other internal mechanisms are the 
most important sources of views for Youtube videos.
Our work is complementary to these as we have also 
sought to understand factors influencing video popularity, but
unlike these prior works we are able to study the significance
and impact of content-agnostic factors while controlling for
differences in video content.

Cha et al.~\cite{Moon:09} noted the presence of YouTube video clones
which they referred to as ``aliases''. 
They observe that aliases tends to ``dilute''
popularity,
as the views for the same content are 
spread out over several videos. 
The authors did not use aliases to study how different factors
influence content popularity, as we have done in our work.

\section{Conclusions}\label{sec:conclusions}

Video sharing services provide a convenient platform for widespread dissemination of content. 
Every day, millions of videos are uploaded and there
are billions of videos viewed for YouTube alone.
Over time, some videos reach iconic status,
while many others are simply forgotten. 

 Our first contribution is a methodology that is able to accurately assess
the impact various content-agnostic factors have on  popularity. 
We identify and collect a large dataset that consists of multiple near-identical copies (called clones) 
of a range of different content; we make this dataset available to the research community.
We then develop a rigorous analysis framework, which allows us to control bias introduced when studying videos that
do not have the same content. 

Using our clone-based methodology, we provide several findings. First, we
show that  inaccurate conclusions may be drawn when 
not controlling for video content. Second, controlling for video content, we observe 
scale-free rich-get-richer behavior, with view count being the most 
important factor except for very young videos.
Third, we find that while the total view count is the strongest predictor, other 
content-agnostic factors can help explain various other aspects of the 
popularity dynamics. 
For example, the uploader's social network can 
be a good predictor for newly uploaded videos.
Finally, we present concrete 
evidence of the first-mover advantage where 
early uploaders
have an edge over later uploaders of the same content. 

While this paper considers video popularity, we note that
     the methodology presented and employed may be applicable to
     other domains and research questions.  For example, a
     popular news story will result in many ``clones'' disseminated
     through different forums.  The relative popularity of each
     such clone will depend on content-agnostic factors such as age,
     previous popularity, and publisher.  Text-based content raises
     some interesting new methodological issues, as it may permit
     a more objective definition of ``clone''.  Such studies are
     left for future work.

\section{Acknowledgements}

The authors are grateful to Shahriar Kaisar for his work on the data collection.  
This work was supported by National ICT Australia (NICTA), 
the Natural Sciences and Engineering Research Council (NSERC) of Canada, 
and CENIIT at Link{\"o}ping University.

\bibliographystyle{abbrv}
{\scriptsize
\bibliography{references}
}

\newpage

\end{document}